\documentclass[12pt]{article}

\usepackage[utf8]{inputenc}
\usepackage[english]{babel}
\usepackage{amsthm}
\usepackage{amsmath}
\usepackage{amssymb}
\usepackage{color}
\usepackage{graphicx}
\usepackage{hyperref}

\def\forb{{\mathrm{forb}}}
\def\Av{{\mathrm{Avoid}}}

\newcommand{\Qed}{\hbox{ }\hfill\rule{2.5mm}{3mm}\medskip}

\def\cX{{\mathcal{X}}}

\def\FF{{\mathbb{F}}}

\def\NN{{\mathbb{N}}}

\def\I{{\mathcal{I}}}
\def\T{{\mathcal{T}}}
\def\0{{\bf 0}}
\def\1{{\bf 1}}
\newcommand{\ncols}[1]{\|{#1}\|}

\newcommand{\prf}[1]{Proposition~\ref{#1}}

\newcommand{\df}[1]{{\bf{#1}}}

\theoremstyle{definition}
\newtheorem{thm}{Theorem}[section]
\newtheorem{lemma}[thm]{Lemma}

\newtheorem{prop}[thm]{Proposition}

\newtheorem{conj}[thm]{Conjecture}
\newtheorem{defn}[thm]{Definition}
\newtheorem{remark}[thm]{Remark}

\title{Forbidden Configurations: Finding the number predicted by the Anstee-Sali Conjecture is NP-hard}
\author{Miguel Raggi}

\begin{document}
\maketitle

\begin{abstract}
Let $F$ be a hypergraph and let $\mathrm{forb}(m,F)$ denote the maximum number of edges a hypergraph with $m$ vertices can have if it doesn't contain  $F$ as a subhypergraph. A conjecture of Anstee and Sali predicts the asymptotic behaviour of $\mathrm{forb}(m,F)$ for fixed $F$. In this paper we prove that even finding this predicted asymptotic behaviour is an NP-hard problem, meaning that if the Anstee-Sali conjecture were true, finding the asymptotics of $\mathrm{forb}(m,F)$ would be NP-hard.
\vskip 5pt
{\small Mathematics Subject Classification: 05D05} 
\end{abstract}

\section{Introduction}
The paper considers an extremal problem in hypergraph theory that results as a natural generalization of Turáns famous problem. 

Some of the most  celebrated extremal results are those of Erd\H{o}s and Stone \cite{ES} and Erd\H{o}s and Simonovits \cite{Simon}. They consider the following problem: Given $m\in\NN$ and a graph $F$, find the maximum number of edges in a graph on $m$ vertices that avoids having a subgraph isomorphic to $F$. 

There are a number of ways to generalize this to hypergraphs. A $k$-uniform hypergraph is one in which each edge has size $k$. Some view $k$-uniform hypergraphs as the most natural generalization of a graph (a graph is a 2-uniform hypergraph) and one might also generalize the forbidden subgraph problem to a forbidden $k$-uniform subhypergraph problem. There are both asymptotic and exact results (e.g. \cite{dCF}, \cite{Pikhurko08}, \cite{F91}). 

Forbidden Configurations is a different (but also natural) generalization that is studied mainly by Richard Anstee and his colaborators. We consider the following problem: Given $m\in\NN$ and a hypergraph $F$, find the maximum number of edges in a simple hypergraph $H$ on $m$ vertices that avoids having a subhypergraph isomorphic to $F$. Surveys on the topic can be found in \cite{survey} and \cite{mythesis}.

We find it convenient to use the language of matrices to describe hypergraphs: Each column of a $\{0,1\}$-matrix can be viewed as an incidence vector on the set of rows.

\begin{defn}Define a matrix to be \df{simple} if it is a $\{0,1\}$-matrix with no repeated columns. Furthermore, if $\alpha$ is a column and $A$ a matrix, define $\lambda(\alpha, A)$ to be the multiplicity of $\alpha$ in $A$. \end{defn}

Note that an $m\times n$ simple matrix corresponds to a \df{simple hypergraph} (or \df{set system}) on $m$ vertices with $n$ distinct edges, where we allow the ``empty edge''. 

\begin{defn}
When $A$ is a $\{0,1\}$-matrix, we denote by $\ncols{A}$ the number of columns in $A$ (which is the cardinality of the associated set system).
\end{defn}

\begin{defn}
 Let $A$ and $B$ be $\{0,1\}$-matrices with the same number of rows. Define the \df{concatenation} $[A|B]$ to be the configuration that results from taking all columns of $A$ together with all columns of $B$. For $t \in \mathbb{N}$, we define the product
    $$t\cdot A := [\underbrace{\ A\ |\ A\ |\ \cdots \ |\ A\ }_{t \text{ times}}].$$
\end{defn}

Our objects of study are $\{0,1\}$-matrices with row and column order information stripped from them. 

\begin{defn}
Two $\{0,1\}$-matrices are said to be \df{equivalent} if one is a row and column permutation of the other. This defines an equivalence relation. An equivalence class is called a \df{configuration}. 
\end{defn}

Abusing notation, we will commonly use matrices (representatives) and their corresponding configurations interchangeably. 

\begin{defn}
For a configuration $F$ and a $\{0,1\}$-matrix $A$ (or a configuration $A$), we say that $F$ is a \df{subconfiguration} of $A$, and write $F \prec A$ if there is a representative of $F$ which is a submatrix of $A$. We say $A$ \df{has no configuration} $F$ (or \df{doesn't contain $F$ as a configuration}) if $F$ is not a subconfiguration of $A$. Let $\Av(m,F)$ denote the set of all simple matrices on $m$-rows with no subconfiguration $F$.
\end{defn}

Our main extremal problem is to compute
  $$\forb(m,F)=\max_{A}\{\ncols{A}\ :\ A\in\Av(m,F)\}.$$

Perhaps some examples are useful:
\begin{itemize}
  \item $\forb(m,\begin{bmatrix}1\\1\end{bmatrix}) = m+1$, since we can take all columns with at most one 1.
  \item $\forb(m,\begin{bmatrix}1\\0\end{bmatrix}) = 2$, since we may only take the column of 1's and the column of 0's (\textit{i.e.} the empty set and the complete set).
  \item $\forb(m,\begin{bmatrix}1 & 1 \\ 1 & 1\end{bmatrix}) = \binom{m}{2}+\binom{m}{1}+\binom{m}{0}$, by taking all columns with at most two 1's. The proof that this is indeed the maximum is easy and can be found in \cite{mythesis}.
\end{itemize}

Let $A^c$ denote the $\{0,1\}$-complement of $A$ (replace every 0 in $A$ by a 1 and every 1 by a 0). Note that $\forb(m,F) = \forb(m,F^c)$.

\begin{remark}\label{inc}
Let $F$ and $G$ be configurations such that $F \prec G$. Then $\forb(m,F) \leq \forb(m,G)$. 
\end{remark}

We say a column $\alpha$ has \df{column sum} $t$ if it has exactly $t$ ones. Let $\0_m$ denote the column with $m$ rows, all of them zeros. Similarly, let $\1_m$ denote the column of $m$ ones. 

For a set of rows $S$, we let $A|_S$ denote the submatrix of $A$ given by restricting the rows of $A$ to only those in $S$.

An important general result due to F\"uredi applies to simple or to non-simple configurations.
\begin{thm}[Z. F{\"{u}}redi]\label{tK_k} Let $F$ be a given $k$-rowed $\{0,1\}-$matrix. Then $\forb(m,F)$ is in $O(m^k)$.
\end{thm}

We desire more accurate asymptotic bounds. Anstee and Sali conjectured that the best asymptotic bounds can be achieved with certain \emph{product constructions}.

\begin{defn}
Let $A$ and $B$ be $\{0,1\}$-matrices. We define the \df{product} $A\times B$  by taking each column of $A$ and putting it on top of every column of $B$. Here is an example of a product:
$$A = \begin{bmatrix}
   0 & 1 & 1 \\
   0 & 0 & 1 \\
  \end{bmatrix}, \ 
  B = \begin{bmatrix}
  1 & 0 \\
  0 & 1 \\
  \end{bmatrix}\ \implies\ \begin{matrix}A\\ \times \\ B\end{matrix} = 
      \left[\begin{array}{cc|cc|cc}
        0 & 0 & 1 & 1 & 1 & 1 \\
        0 & 0 & 0 & 0 & 1 & 1 \\
        \hline
        1 & 0 & 1 & 0 & 1 & 0 \\
        0 & 1 & 0 & 1 & 0 & 1 \\
      \end{array}\right].$$
      
      Note that this is a well defined operation in configurations.
\end{defn}

We are interested in asymptotic bounds for $\forb(m,F)$. Let $\I_m$ be the $m\times m$ identity matrix, $\I_m^c$ be the $\{0,1\}$-complement of $\I_m$ (all ones except for the diagonal) and let $\T_m$ be the tower matrix: a matrix corresponding to a maximum chain in the partially ordered set of the power set of the vertices. For example, 
$$\T_4 = \begin{bmatrix}
		    0 & 1 & 1 & 1 & 1\\
		    0 & 0 & 1 & 1 & 1\\
		    0 & 0 & 0 & 1 & 1\\
		    0 & 0 & 0 & 0 & 1\\
		  \end{bmatrix}$$

Anstee and Sali conjectured that the asymptotically ``best''  constructions avoiding a single configuration would be products of $\I,\I^c$ and $\T$.

\begin{defn}
  Let $F$ be a configuration. Let
  $$P_r(a,b,c) := \underbrace{\I_r\times ... \times \I_r}_{a\text{ times}} \times \underbrace{\I_r^c \times ... \times \I_r^c}_{b\text{ times}} \times \underbrace{\T_r \times ... \times \T_r}_{c \text{ times}},$$

Define $X(F)$ to be the largest number such that there exist numbers $a,b,c \in \NN$ with $a+b+c = X(F)$ such that for all $r \in \NN$,
 $$F \nprec P_r(a,b,c).$$
\end{defn}

\begin{conj}\label{grand}\cite{AS05}
  Let $F$ be a configuration. Then $\forb(m,F)$ is in $\Theta(m^{X(F)}).$
\end{conj}

Observe that $X(F)$ is always an integer. Also note that $\ncols{P_r(a,b,c)} = r^{a+b}\cdot (r+1)^c \in \Theta(r^{X(F)})$, so by taking $r = \lceil m/X(F) \rceil$ (and perhaps deleting some rows in case $X(F)\nmid m$), we have that $\ncols{P_r(a,b,c)} \in \Omega(m^{X(F)})$. So the fact that $\forb(m,F)\in \Omega(m^{X(F)})$ is built into the conjecture. 

Thus, in order to prove the conjecture, all that would be required would be to prove that $\forb(m,F) \in O(m^{X(F)})$ for every $F$. A disproof could be potentially easier, as only a counterexample would be required.

The conjecture has been proven for all $k\times \ell$ configurations $F$ with $k \in \{1,2,3\}$ and many others cases in various papers. The proofs for $k=2$ are in \cite{AGS}, for $k=3$ in \cite{AGS}, \cite{AFS}, \cite{AS05}. For $\ell = 2$, the conjecture was verified in \cite{AK}. For $k=4$, all cases either when the conjecture predicts a cubic bound for $F$ or when $F$ is simple were completed in \cite{AF}. For $k=4$ and $F$ non-simple, there are three boundary cases with quadratic bounds, one of which is established in \cite{F8}. For $k\in\{5,6\}$ some results can be found in \cite{F7}.

Anstee has long conjectured that even finding $X(F)$ given $F$ was not a trivial task, and the question of its NP-hardness was long conjectured (\cite{survey}, \cite{mythesis}). In this paper we settle this question: finding $X$ is indeed NP-hard. We also note that one of the decision versions associated with this optimization problem is NP-complete, adding this function to the long list of functions known to be NP-complete, with the interesting plus that this function is conjectured to give the \emph{exponent} of the asymptotic growth of $\forb$.

For relatively small configurations $F$ we have a computer program that yields the answer (relatively) quickly. The source code (in C++) can be freely downloaded from:
	\begin{center}
	\url{http://matmor.unam.mx/~mraggi/}
	\end{center}

This program can compute $X(F)$ for $F$ having less than $\sim$10 rows in just a few minutes. This task takes merely exponential time, not doubly exponential (as it is often the case with forbidden configuration problems). This program was written to perform many other tasks other than finding $X(F)$. A description of the algorithm used for this task is in Section \ref{appendix}.

\section{Results}

There are two natural decision problems associated with $X(F)$: Given $F$ and $k$ as inputs,
\begin{enumerate}
  \item Is it true that $X(F) < k$?
  \item Is it true that $X(F) \geq k$?
\end{enumerate}

We prove that the first of the two decision problems is in NP by exhibiting a certificate which can be checked in polynomial time.

The main result of this paper is the following:
\begin{thm}\label{np}
  Finding $X(F)$ is an NP-hard problem. In other words, should a polynomial-time algorithm exist for finding $X(F)$ given $F$, then every problem in NP could be solved in polynomial time. Furthermore, the problem ``given $F$ and $k$, is $X(F) < k$?'' is in NP.
\end{thm}

Before proving this theorem, we need a few lemmas.

\begin{lemma}\label{nomasrenglones}
  Let $F$ be a configuration with $n$ rows. Then $X(F) \leq n$. 
\end{lemma}

\textbf{Proof}: Indeed, assume for the sake of contradiction that $a$, $b$, and $c$ are such that $a+b+c=n+1$ and $F \nprec P_r(a,b,c)$. We may place each row of $F$ each into a different factor of the product. The ``extra matrix'' ensures the columns of $F$ with high multiplicity get repeated as many times as needed. \Qed

This observation in particular implies that if a polynomial time algorithm existed for any of the two decision versions of the problem, then we'd have a polynomial time algorithm for finding $X(F)$, which together with Theorem \ref{np} would make the decision version of finding $X(F)$ an NP-complete problem.

\vspace{0.3cm}

A simple (but surprising) corollary of Conjecture \ref{grand}, if it were true, would be that repeating columns more than twice in $F$ has no effect on the asymptotic behavior of $\forb(m,F)$. In other words, assuming the conjecture were true, the multiplicity of a column in a configuration would not affect the asymptotic bound and, asymptotically, it would only matter if a column is not there (has multiplicity 0), appears once (has multiplicity 1), or appears ``multiple times'' (has multiplicity 2 or more). Formally,

\begin{prop}\label{cortoconj}
 Let $F_t=[G|t\cdot H]$ with $G$ and $H$ simple $\{0,1\}$-matrices that have no columns in common. Then $X(F_2) = X(F_t)$ for all $t \geq 2$. In particular, if the conjecture were true, then $\forb(m,F_t)$ and $\forb(m,F_2)$ would have the same asymptotic behavior.
\end{prop}

\textbf{Proof}: It suffices to show that given $t$, $G$, $H$, $a$, $b$ and $c$, there exists an $R$ such that for every $r \geq R$, we have
  $$F_2 = [G|2\cdot H] \prec P_r(a,b,c)\quad
  \iff\quad
  F_t = [G|t\cdot H] \prec P_r(a,b,c). $$

Since $F_2 \prec F_t$, we only need to prove that if $F_2 \prec P_r(a,b,c)$ for some $r$, then $F_t \prec P_R(a,b,c)$ for some $R$. Suppose then $F_2$ is contained in the product $P_r(a,b,c)$ for some $r$. The idea is to find a subconfiguration of $P_r(a,b,c)$ in which there are some columns with multiplicity 1, and for the columns with multiplicity 2 or more, the multiplicity depends on $r$. We need $r$ large enough so that the multiplicity of any one column (with multiplicity of 2 or more) is larger than $t$. Let $x$ be the number of rows of $F_t$. Notice the following three facts, which include definitions for $E_\I$, $E_{\I^c}$ and $E_\T$.
  \begin{align*}
  E_\I(x,r) := [(r-x)\cdot\0_{x}\ |\ \I_x ] & \prec \I_r \\
  E_{\I^c}(x,r) :=  [(r-x)\cdot\1_{x}\ |\ \I^c_x ] & \prec \I^c_r \\
  E_{\T}(x,r) :=  \left\lfloor \frac{r}{x}\right\rfloor \cdot \T_x & \prec \T_r.\end{align*}

The first and second facts are easy to see; just take any subset of $x$ rows from $\I_r$ or $\I^c_r$. The third statement is true by taking the $\lfloor r/x\rfloor$-th row of $\T_r$, the $2\lfloor r/x\rfloor$-th row of $\T_r$, etcetera, up to the $x\lfloor r/x\rfloor$-th row. For example, if $r = 5$ and $x = 2$, we may take the second and fourth row from $\T_5$:
    $$\T_5 = \begin{bmatrix}
             0 & 1 & 1 & 1 & 1 & 1\\
             0 & 0 & 1 & 1 & 1 & 1\\
             0 & 0 & 0 & 1 & 1 & 1\\
             0 & 0 & 0 & 0 & 1 & 1\\
             0 & 0 & 0 & 0 & 0 & 1\\
            \end{bmatrix}\ \ 
\implies\ \ \T_5|_{\{2,4\}} =
\begin{bmatrix}
0 & 0 & 1 & 1 & 1 & 1\\
0 & 0 & 0 & 0 & 1 & 1\\
\end{bmatrix} = E_\T(2,5)$$

Note that in the three configurations $E_\I(x,r), E_{\I^c}(x,r)$ and $E_\T(x,r)$, we have that there are some columns of multiplicity 1 and there are some columns for which their multiplicity can be made as large as we wish by making $r$ large. Formally, let $E(x,r)$ be one of $E_\I(x,r)$ or $E_{\I^c}(x,r)$ or $E_\T(x,r)$. We have that for every $x$-rowed column $\alpha$ there are three possibilities: either $\lambda(\alpha, E(x,r)) = 0$ for all $r$, or $\lambda(\alpha, E(x,r)) = 1$ for all $r$, or $\displaystyle\lim_{r\to \infty} \lambda(\alpha, E(x,r)) = \infty.$

If $\alpha$ is a column for which $\displaystyle\lim_{r\to \infty} \lambda(\alpha, E(x,r)) = \infty$, we may conclude that there is an $R$ for which $\lambda(\alpha,E(x,r)) \geq t$ for every $r \geq R$.

Since $F_2$ is contained in $P_r(a,b,c)$ for some $r$, the columns in $H$ will have multiplicity at least 2 in some subset of the rows of $P_r(a,b,c)$. We see that $F_t$ is also a subconfiguration of $P_R(a,b,c)$.\Qed

\section{Proof of the Main Theorem}
We now prove the main theorem. 

\textbf{Proof}: First we prove that the decision problem has a certificate which can be checked in polynomial time. A certificate that indeed $X(F) < k$ would have to be a proof that $F \prec P_r(a,b,c)$ for each triple $a,b,c$ for which $a+b+c=k$. Note that there are at most a quadratic (with respect to the number of rows) number of $a,b,c$'s which satisfy the equation, since the question has a trivial ``yes'' answer when $k$ is more than the number of rows (Lemma \ref{nomasrenglones}). 

Given $F$ and $A$ configurations, one can easily construct a certificate that a configuration $F$ is indeed a subconfiguration of a configuration $A$: explicitly state which permutation of $F$ appears in exactly which rows and columns of $A$. For the case $A=P_r(a,b,c)$, a certificate only needs to specify which rows of $F$ go inside which factors, so at most a quadratic number of these certificates-of-being-a-subconfiguration suffice.

We now prove that finding $X(F)$ is NP-hard. Suppose there existed some polynomial-time algorithm that finds $X(F)$ given $F$. We shall prove that there would then exist a polynomial time algorithm for GRAPH COLORING. Suppose we are given a graph $G$ and we wish to find the minimum number of colors for which there exists a good coloring of the graph. We may assume no isolated vertices.

The idea is to construct a 3-part matrix $F(G)$ in which the first two parts ensure there is no $\T$ or $\I^c$ in a maximum product of the form $P_r(a,b,c)$ with no subconfiguration $F(G)$, and the last part is constructed so that a partition into $\I's$ produces a partition of the vertices of the graph into independent sets and vice-versa.

Suppose $G$ has $n$ vertices and $e$ edges. Let $M$ be a large number with $M \geq n+2$ and let $S$ be the incidence matrix of $G$ (\textit{i.e.}, the edges of $G$ are encoded as columns with two 1's corresponding to the vertices that belong to the edge). Construct the following simple matrix:
	$$F(G) := \left[\begin{array}{c|c}
	    1 & \0_M \I^c_M \\
	    \hline
	    1 & \T_M \\
	    \hline
	    S & 0 \\
	  \end{array}\right]
	$$

Clearly we can construct $F(G)$ in polynomial time (with respect to the number of vertices of $G$). We prove now that we have $\chi(G) = X(F) - 2M + 1$, which in turn would yield a polynomial time algorithm for GRAPH COLORING, provided we had a polynomial time algorithm for $X(F)$.

Now, let us study the possibilities for a product of type $P_r(a,b,c)$ that does not have $F(G)$ as a subconfiguration for any $r$. If $b\neq 0$, then we could place all of $[1 | \I^c_M]$ in the $\I^c$ part of $P_r(a,b,c)$, so $a+b+c$ would be at most $1+M+n$ (using Lemma~\ref{nomasrenglones}). The same is true when $c\neq 0$. But if we let $b=c=0$, $P_r(a,0,0)$ is just a product of $\I$'s, so let us calculate how many $\I's$ we can multiply together and still not create a subconfiguration $F(G)$. 

In order for $F(G)$ to be a part of a product of $\I's$, every row of $[1 | \I^c_M]$ and $[1 | \T_M]$ must be in a separate factor $\I$, since there is no $\begin{bmatrix}1\\1\end{bmatrix}$ in $\I$ (and also separate from the rows of $[S | 0]$, since we are assuming $G$ has no isolated vertices).

Then two rows of the $[S | 0]$ part can be in the same $\I$ if and only if there is no $\begin{bmatrix}1\\1\end{bmatrix}$ in those two rows, which, in terms of the graph, means there is no edge between those two vertices. In other words, partitioning $[S | 0]$ into $I$'s is equivalent to partitioning the vertices of $G$ into independent sets. So if the graph $G$ cannot be colored with $\chi(G)-1$ colors and this is the maximum, this means that $X(F) = a = 2M + \chi(G) - 1 \geq n+M+1$. Then $\chi(G) = X(F(G)) - 2M + 1$. \Qed

\section{Algorithm to find X(F)}\label{appendix}

In this section we describe the algorithm used by the software described in the introduction. It runs in exponential time, of course, but it has been helpful for finding \emph{boundary configurations} with given asymptotic bounds (see \cite{F7}, \cite{mythesis} and \cite{survey}). Perhaps this program might be used to find a counter example to the Anstee-Sali conjecture, provided one exists, and it isn't very large.

\subsection{Representation of a Configuration}
  We are interested in an efficient representation for configurations in order to perform the tasks described above. In the progress of our investigations, we have had various versions of the program. 
  
  Most of what we want the program to do involves performing a huge number of \emph{configuration comparison operations}, which is testing whether or not a configuration $F$ is a subconfiguration of a configuration $A$. As a first approach it would seem as if, for this task, we would be required to test each row and column permutation of $F$ against each submatrix of $A$. This is of course a very slow way to do this. A simple trick to speed up the computations is to keep the columns of a configuration always in some canonical order. Then, to test whether or not a configuration $F$ is contained in another $A$, we just need to permute rows of $F$ and take subsets $S$ of rows of $A$ and place the columns of $A|_S$ in canonical order.
  
  Most of the tasks we wish this program to perform involve checking whether or not a given (fixed) configuration $F$ is a subconfiguration of a vast number of configurations $A$. In particular, any pre-processing we do on $F$ can be considered as almost free. For example, finding all row permutations of $F$ and storing them would need to be done once for each configuration $F$, and not at all for configurations $A$.
  
  After many attempts and experiments, it seemed that the best (fastest) way to store a configuration that made many of the other tasks reasonably fast is this: Maintain an array of integers where the \emph{indices} of the array, written in binary, are the \emph{columns} of the configuration, and the actual numbers of the array represent the number of times a column appears. That is to say, a configuration $F$ in $m$ rows is represented by an array (C++ vector) $\FF$ of size $2^m$. For a number $\alpha$, consider the binary representation of $\alpha$ and consider it as a column with $m$ rows. If necessary, put enough 0's at the beginning of the binary representation in order to have the required $m$ bits. The number $\FF[\alpha]$ (the $\alpha$-th number of the array) represents the number of times that column $\alpha$ appears in configuration $F$. In the implementation, we use an array of \emph{unsigned characters} instead of integers, since we never need a configuration with the same column repeated more 
than 255 times. An unsigned character consists of 1 byte (8 bits).
  
  For example, the array $\FF = [1,0,0,2,0,1,0,1]$ represents the following configuration (notice it has 3 rows, since the array has size $8 = 2^3$):
      $$F = \begin{bmatrix}
          0 & 0 & 0 & 1 & 1 \\
          0 & 1 & 1 & 0 & 1 \\
          0 & 1 & 1 & 1 & 1 \\
        \end{bmatrix}.$$
  
  To see this, remember we start from 0. There is a one in position $0=000_b$, meaning the colum $(0,0,0)^T$ gets repeated one time. A two in position $3 = 011_b$, a one in position $5=101_b$ and a one in position $7=111_b$. The columns of this matrix are the representations of these numbers in binary form.
  
  An observant reader might complain that this has the disadvantage that it requires storing $2^m$ bytes, and if $F$ doesn't have many columns, most of those will be 0's. But it's a minor disadvantage, because even at 10 rows we would only need 1024 bytes, and we usually have configurations for which the number of rows is 5 or less (32 bytes). Perhaps this would become more of an issue with configurations with a high number of rows, but for those configurations, most of our tasks would require too much time to be of any practical use anyway.
  
  We came to this representation after an implementation which represented columns as an array of bits (C++ bitset) and storing them into an ordered tree-like structure (C++ multiset from the STL). This might be a more natural implementation, but profiling the code made clear that the program was spending most of its time counting how many columns of a certain type appeared in a configuration, and was also spending a considerable amount of time navigating the tree. Explicitly storing the number of times each column appears, and making that number instantly accessible by storing it in an array (for random access) gives a very noticeable speedup and allows us to consider larger problems. By representing columns as numbers, we can do a lot of preprocessing and compute large tables in which we have almost instant access time. 
  
  For example, consider the following problem, which has to be done many times for our tasks: Given a column $\alpha$ and a subset $S$ (represented by an integer as written in binary), what column is $\alpha_S$? This is relatively slow to compute, but we can fill out a table by preprocessing to speed up any further access to it. Since we do this a few million times, the investment is sound.
  
  The other advantage is that it becomes immediately clear how to compare two configurations with the same number of rows to see if one is a column-permutation submatrix of the other; check if for any column (index) the integer at position $c$ of the first array is bigger than that of the second array. To check if $F$ is a subconfiguration of $A$, we would need to find all permutations of the rows of $F$ (which we need to do just once per configuration).
  
  \subsection{Subconfigurations}
  Suppose $F$ and $A$ are configurations and we want to decide if $F \prec A$. Then for every $s$-tuple of rows $S$ (where $s$ is the number of rows of $F$), we can extract from $A$ the configuration $A|_S$ easily with our pre-stored table of columns and subsets. Once we've done this for each column of $A$ and found $A|_S$, then for each permutation of rows of $F$, we check if every column $\alpha$ in the array corresponding to configuration $F$ appears less than or equal to the corresponding number for column $\alpha$ in the array of $A|_S$. We can check every subset $S$ like this. If at any point this is so, we can return true.
  
  There are a few speedups. Sometimes it's immediately obvious a configuration can't be contained in another. For example, if there are more 1's in $F$ than in $A$, or if $F$ has more columns or rows than $A$, then $F \nprec A$.

  \subsection{Determining X(F)}\label{sec:conjcomp}
  Given a configuration $F$, we wish to find $X(F)$. In other words, we wish to find the conjectured asymptotic bound for $\forb(m,F)$.
  
  We may make a simplification using \prf{cortoconj} and assume the multiplicity of any column of $F$ is at most 2.
  
  First, suppose we wanted to test whether or not there exists $r$ such that configuration $F$ is contained in the product 
    $$P_r(a,b,c) = \underbrace{\I_r\times ... \times \I_r}_{a\text{ times}} \times \underbrace{\I_r^c \times ... \times \I_r^c}_{b\text{ times}} \times \underbrace{\T_r \times ... \times \T_r}_{c \text{ times}}.$$
  
  Building this object with $r=R$ as calculated in \prf{cortoconj} would be prohibitively slow. Instead, we build a set $\cX$ of subconfigurations from $P_R(a,b,c)$ such that if $F\prec P_R(a,b,c)$, then $F \prec X$ for some $X\in\cX$. 
  
  Recall that if $F\prec P_R(a,b,c)$, then the rows of $F$ get partitioned into $a+b+c$ parts (a part can be empty), where each part belongs to a factor of the product $P_R(a,b,c)$. Because of \prf{cortoconj}, we can assume each column appears at most twice. Given $s \in \NN$, consider the following matrices:
  $$A_\I(s) = \begin{bmatrix}
             t\cdot 0_s & \I_s \\
            \end{bmatrix}\ ,\quad
  A_{\I^c}(s) = \begin{bmatrix}
             t \cdot 1_s & \I^c_s \\
            \end{bmatrix}\ ,\quad
  A_\T(s) = t\cdot \begin{bmatrix}
             \T_s \\
            \end{bmatrix}.$$
  
  We see that an $s$-rowed configuration $F$ (with each column repeated at most $t$ times) is contained in $I_m$ for some large $m$, if and only if $F \prec A_I(s)$. We can then consider all partitions of rows of $F$ and see if each part is contained in the corresponding $A_\I$, $A_{\I^c}$ or $A_\T$.
  
  For example, to test whether
    $$F = \begin{bmatrix}
           1 & 0 & 1 & 1 \\
           0 & 1 & 0 & 1 \\
           1 & 1 & 0 & 0 \\
           0 & 0 & 1 & 1 \\
           1 & 0 & 0 & 1 \\
          \end{bmatrix}$$
is contained in $\I \times \T \times \T$, we would partition the rows of $F$ in three parts. In this case, $F$ has five rows, so consider, for example, the following partition of 5: $(2,2,1)$. Consider the following representatives:
  $$A_{\I}(2) = \left[ s \cdot \begin{bmatrix}0 \\ 0\end{bmatrix}\ \left|\ \begin{bmatrix} 1 & 0 \\ 0 & 1 \end{bmatrix}\right]\right. , \quad
    A_{\T}(2) = s\cdot \begin{bmatrix}0 & 1 & 1 \\ 0 & 0 & 1\end{bmatrix}, \quad
    A_{\T}(1) = s\cdot \begin{bmatrix} 0 & 1 \end{bmatrix},$$
and build a 5-rowed matrix $A := A_{\I}(2) \times A_{\T}(2) \times A_{\T}(1)$. If $F \prec A$, then $F \prec \I \times \T \times \T$. If we do this for every possible partition of $5$, we get the desired result.

To find $X(F)$ we can build a tree of possibilities of products and prune whenever we hit a node that already has configuration $F$: 
\begin{center}
    \includegraphics[scale=0.8]{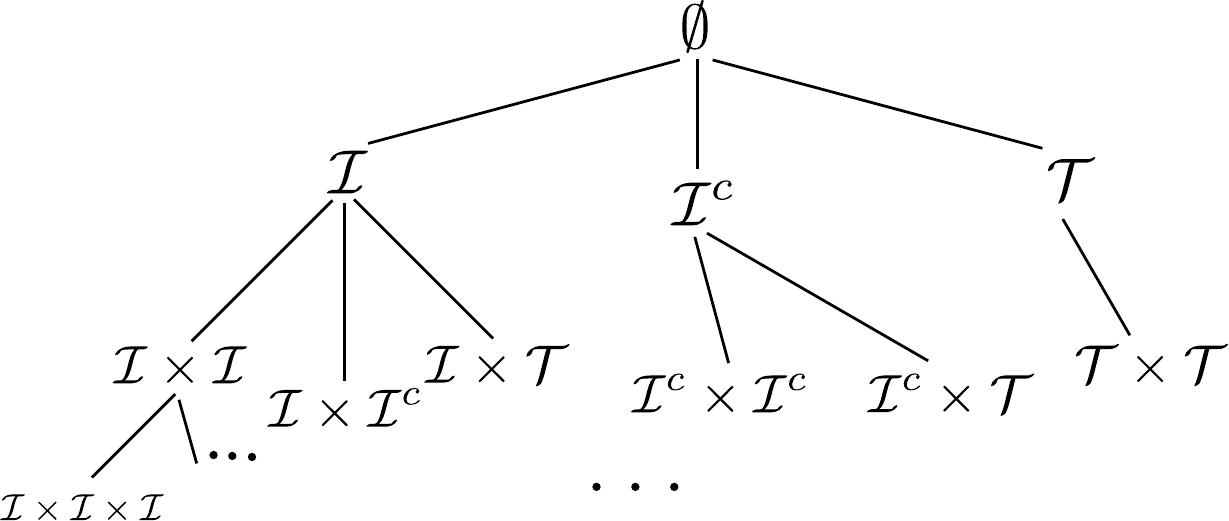}
\end{center}

We return the product with the largest number of factors for which $F$ is not a subconfiguration of a product of this form, observing that if $F$ is a subconfiguration of a product $P_r(a,b,c)$, then it will be a subconfiguration in any product $P_r(a',b',c')$ with $a'\geq a$, $b'\geq b$ and $c'\geq c$.

\subsection{Finding Boundary Cases and Classifying Configurations}\label{sec:classifycomp}
  Given number of rows $s$ and a number $k$, we wish to find all \emph{boundary} cases, that is maximal and minimal $F$ such that $X(F) = k$. We make use of the program described in the previous section to find $X(F)$ for many different $F$. The boundary cases for small $s$ and $k$ can be found in \cite{survey} and \cite{mythesis}.
  
  The method we use is very straightforward: start adding columns, one by one. Build a tree of configurations, where the children of a configuration $F$ are the ones that consist of $F$ plus a column. Find $X(F)$ for each configuration in the tree. Store all those configurations $F$ for which $X(F) = k$, and then find only the maximal and minimal configurations in the ordering $\prec$. If at some point we add a column and the bound jumps to $k+1$ or higher, discard and go to the next configuration. Because of \prf{cortoconj}, we only need to consider cases where columns are repeated at most twice.

  Unfortunately this method is very slow because the same configuration is searched multiple times, since each configuration may have many representatives. To get rid of repetition we might check for equivalence of configurations against everything we have stored so far. But checking if two configurations are equivalent is usually slow, so doing it every time is also very slow.
  
  What seems to work best is to do check for equivalence, but only up to a point. For example, we can consider all pairs of columns, test those and only take one representative of each equivalence class. Then from each pair, start building the tree as described above. After that, it will be relatively unlikely that two configurations we search are equivalent, so the amount of repetition will be relatively low. Of course, much repetition will still occur, but much less than in the original tree.
  
  Notice that it isn't as critical that classifying configurations be a fast calculation. We need to do it once for every $s$ and $k$, but no more. Once we know the maximal and minimal quadratics for five rows, we never need to calculate them again. Other calculations, such as finding $X(F)$ or What Is Missing, have to be performed for each configuration (or family) we wish to study, so decreasing the running time is more of a priority in those cases since faster programs allow us to study bigger configurations.

\bibliographystyle{amsalpha}
\bibliography{FCsurvey10}

\end{document}